# Randomness Quantification for Quantum Random Number Generation Based on Detection of Amplified Spontaneous Emission Noise

Jie Yang[1], Fan Fan[1], Jinlu Liu[1], Qi Su[2], Yang Li[1], Wei Huang[1], and Bingjie Xu[1,*]

[1] Science and Technology on Security Communication Laboratory, Institute of Southwestern Communication, Chengdu 610041, China
[2] State Key Laboratory of Cryptology, Beijing 100878, China

E-mail: xbjpku@163.com

**Abstract**

The amplified spontaneous emission (ASE) noise has been extensively studied and employed to build quantum random number generators (QRNGs). While the previous relative works mainly focus on the realization and verification of the QRNG system, the comprehensive physical model and randomness quantification for the general detection of the ASE noise are still incomplete, which is essential for the quantitative security analysis. In this paper, a systematical physical model for the emission, detection and acquisition of the ASE noise with added electronic noise is developed and verified, based on which the numerical simulations are performed under various setups and the simulation results all significantly fit well with the corresponding experimental data. Then, a randomness quantification method and the corresponding experimentally verifiable approach are proposed and validated, which quantifies the randomness purely resulted from the quantum process and improves the security analysis for the QRNG based on the detection of the ASE noise. The physical model and the randomness quantification method proposed in this paper are of significant feasibility and applicable for the QRNG system with randomness originating from the detection of the photon number with arbitrary distributions.

## 1. Introduction

Random numbers are of extreme importance for a wide range of applications in both commercial and scientific fields [1], such as numerical simulations [2], lottery games and cryptography [3]. Especially, with the rapid development of quantum key distribution (QKD) system, which is the most practical application in quantum information technology, the demand for true random numbers with a generation rate over Gbps is inevitable [4]. Conventional pseudo random number generators (PRNGs) based on computational algorithms can expand a short random seed into a long sequence of binary bits that appears truly random and have been widely used in modern digital electronic information systems. However, due to the deterministic and thus predictable features of the algorithms, PRNGs are not suitable for the applications that require true randomness, for instance, the cryptography and QKD system [5-7].

Distinct from PRNGs, quantum random number generators (QRNGs) are based on the intrinsic randomness of fundamental quantum processes and can provide truly unpredictable and irreproducible random numbers with very high generation rates [8-10]. Over the past two decades, various QRNG schemes have been proposed and demonstrated, including the detection of photon path [11-13], photon arrival times [14-18], photon number distribution [19-22], vacuum fluctuations [23-28], quantum phase fluctuations [29-39] and amplified spontaneous emission (ASE) noise [4, 40-46], etc.

Among all the approaches to build a QRNG, the ASE noise has attracted remarkable attentions and has been widely studied. On one hand, the spontaneous emission is a well understood quantum random phenomenon and the ASE noise is the amplified result of the spontaneous emission with a random intensity which can be easily measured by a photodetector (PD) directly without the need of stable interference control [4, 40-45]. On the other hand, the ASE noise can be easily generated by employing a fiber amplifier or a superluminescent diode (SLED) [4, 40-46]. Besides, the ASE noise usually shows a flat spectrum in a very wide frequency range and therefore can be employed to generate random numbers at very fast rates by using high speed detection and acquisition systems. With the advantages listed above, various QRNG schemes based on the detection of ASE noise have been proposed and demonstrated with off-the-shelf components and high generation rates have been achieved [4, 40-46]. While the previous relative works mainly focus on the realization and verification of the QRNG system, the comprehensive physical model and randomness quantification for the general detection of the ASE noise are still incomplete, which is essential for the quantitative security analysis. Particularly, in [4] it is pointed out that the intensity distribution of the ASE noise can be descried by the Bose-Einstein distribution, but no verification of this theoretical distribution is given and the randomness quantification based on the theoretical distribution is still missing.

In this paper, a systematical physical model for the emission, detection and acquisition of the ASE noise with added electronic noise is developed in detail. By means of experimental validation combined with numerical simulation, the physical model is validated and the statistical distribution of the ASE noise from a SLED in general detection is quantitatively verified. The numerical simulations for the physical model are performed under various setups and the simulation results all significantly fit well with the corresponding experiment data. Then based on the physical model, a randomness quantification method and the corresponding experimentally verifiable approach are proposed and validated, which quantifies the randomness purely resulted from the quantum process and improves the security analysis for the QRNG based on the detection of the ASE noise. The physical model and the randomness quantification method proposed in this paper are of significant feasibility and applicable for the QRNG system with randomness originating from the detection of the photon number with arbitrary distributions.

## 2. Experimental setup and Numerical Simulation

As mentioned above, many QRNG schemes are based on the direct detection of the ASE noise. In this section we focus on the physical model for the general detection of the ASE noise, which explains the physical nature of the detected ASE noise. This is of significant importance because it reveals what is obtained through the detection and where the randomness essentially originates. By means of experimental validation and numerical simulation, the physical model is validated and the photon statistics of the optical field of the ASE noise from a SLED in general detection is quantitatively verified under various setups.

## 2.1 Photon Statistics of the ASE Noise

As stated in [4, 47-49], the photon statistics of the ASE noise in single mode can be described by the Bose-Einstein distribution

$$P_{BE}(n,\bar{n}) = \frac{\bar{n}^n}{(1+\bar{n})^{1+n}} \qquad (1)$$

where $P_{BE}(n,\bar{n})$ stands for the probability of counting $n$ photons by the PD during its average detection time $T$, which is defined as the inverse of the PD bandwidth, and $\bar{n}$ is the average number of photons within the same time interval.

More generally, for the optical field of the ASE noise that contains more independent modes, the photon distribution can be noted by the $M$-fold degenerate Bose-Einstein distribution [4, 47-49]

$$P_{BE}(n,\bar{n},M) = \frac{\Gamma(n+M)}{\Gamma(n+1)\Gamma(M)}\left(1+\frac{1}{\bar{n}}\right)^{-n}(1+\bar{n})^{-M} \qquad (2)$$

where $\Gamma(x)$ is the gamma function, $M$ is the number of independent modes and $\bar{n}$ is the average number of photons of each mode, respectively.

In particular, for the ASE noise with a Gaussian optical spectrum, $M$ can be calculated as [48, 49]

$$M = s \frac{\pi(B_{opt}/B_{ele})^2}{\pi(B_{opt}/B_{ele})\,\text{erf}\left(\sqrt{\pi}(B_{opt}/B_{ele})\right) - \left[1-\exp\left(-\pi(B_{opt}/B_{ele})^2\right)\right]} \qquad (3)$$

where $\text{erf}(x)$ is the error function, $B_{opt}$ is the optical bandwidth, $B_{ele}$ is the electrical bandwidth and $s$ is the polarization degeneracy which equals to 1 (for polarized ASE noise) or 2 (for unpolarized ASE noise), respectively. Therefore, the mode number $M$ can be totally determined by the polarization degeneracy $s$ and the optical-to-electrical bandwidth ratio $B_{opt}/B_{ele}$. A brief simulation is plotted in Fig. 1.

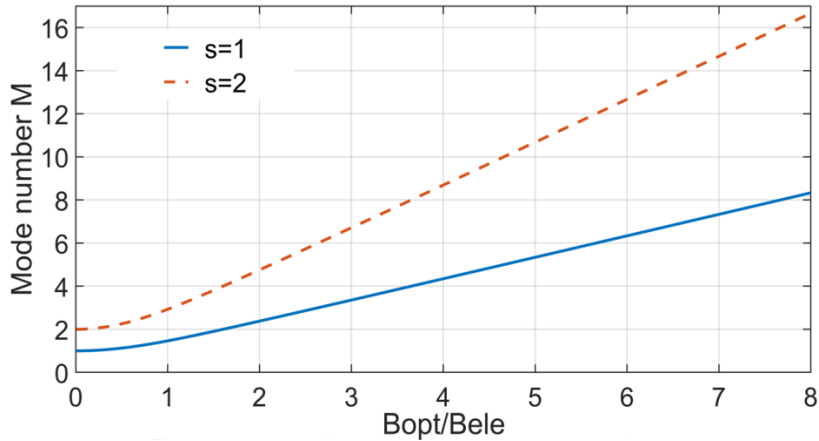

Figure 1. The mode number $M$ as a function of the polarization degeneracy $S$ and the optical-to-electrical bandwidth ratio $B_{opt}/B_{ele}$.

From Fig. 1 we can see that for polarized ASE noise, the mode number $M$ asymptotically tends to the ratio $B_{opt}/B_{ele}$ when the optical bandwidth is obviously larger than the electrical bandwidth whereas $M$ tends to unity for an optical bandwidth much smaller than the electrical bandwidth [48]. For the unpolarized ASE noise, the mode number $M$ is simply double the value of that obtained in the polarized situation.

In practical experiments, it is not difficult to obtain an expected value of mode number $M$ based on the control of the optical-to-electrical bandwidth ratio $B_{opt}/B_{ele}$, which can be properly realized by employing the optical filtering. Meanwhile the average number of photons $\bar{n}$ during the average detection time of PD can also be measured and calculated. Thus, the theoretical distribution of the PD detection result based on equation (2) can be achieved, which will be introduced in the next subsection.

## 2.2 Experimental Setup

The experimental setup is shown in Fig. 2. A SLED (EXSLOS, EXS210059-01) with a typical center wavelength of 1550nm is driven by a butterfly packaged laser diode driver to generate ASE noise. An erbium doped fiber amplifier (EDFA) is employed to enhance the optical power of the ASE noise. Then the ASE noise is filter by a flexible optical filter (Finisar, Waveshaper 100A) featured with its filter shape can be set as gaussian or bandpass and filter bandwidth can be increased from 10 GHz to 1 THz with a resolution of 1 GHz in a very wide operating frequency range, which enables us to obtain an expected gaussian shape optical spectrum with specific bandwidth. After the filtering, the ASE noise passes through a fiber polarizer to obtain the polarized optical signal. Next, the polarized ASE noise is measured by an optical spectrum analyzer (OSA, Yokogawa, AQ6370D) and a power meter (PM, Thorlabs, PM20) to obtain the optical bandwidth $B_{opt}$ and the optical power $P$, which will be used in the calculation of the mode number $M$ and the average photon number $\bar{n}$, respectively. Finally, the polarized ASE noise is detected by a 5 GHz PD (Thorlabs, DET08CFC) and an 8 GHz digital storage oscilloscope (DSO, Keysight, DSOV084A) is employed to acquire the detected signal $V_{det}$. Besides, in every experiment, the electronic noise $V_{ele}$ when turning off the SLED and EDFA is also acquired. The sampling rate of the DSO is fixed at 10G Sa/s and the sample length of $V_{det}$ and $V_{ele}$ are both $10^7$.

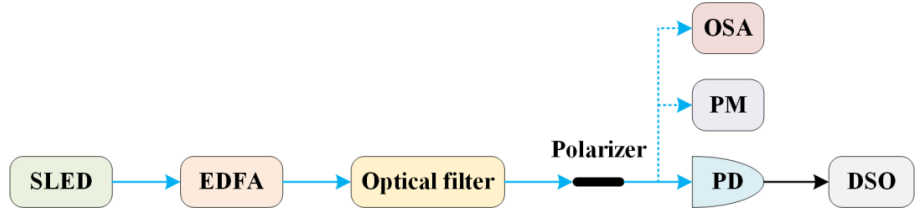

Fig.2. The experimental setup. The ASE noise emitted by the SLED is first amplified by the EDFA. Then the amplified ASE noise is filtered by the optical filter to obtain the expected optical spectrum shape and bandwidth. A fiber polarizer is placed after the optical filter to obtain the polarized optical signal. Next, the polarized ASE noise is measured by an OSA and a PM to obtain the optical spectrum bandwidth and the optical power. Finally, the polarized ASE noise is detected by a PD and then acquired by a DSO to obtain the experimental data. SLED: superluminescent light emitting diode, EDFA: erbium doped fiber amplifier, PD: photodetector, OSA: optical spectrum analyzer, PM: power meter, DSO: digital storage oscilloscope.

Obviously, in our experimental setup, the electrical bandwidth is determined by the PD, i.e., $B_{ele}$ is 5 GHz, and the polarization degeneracy $s=1$. Then with the optical bandwidth $B_{opt}$ measured by the OSA, the mode number $M$ can be easily calculated according to equation (3).

The calculation of the average photon number $\bar{n}$ is also not difficult. Suppose the optical power measured by the PM is $P$, then the average photon number within the detection time can be approximated as

$$\overline{n_M} = \frac{P \cdot T}{h\frac{c}{\lambda}} = \frac{P \cdot 1/B_{ele}}{h\frac{c}{\lambda}} \qquad (4)$$

where $h$ is the Planck constant, $c$ is the velocity of light in vacuum, $\lambda$ is the center wavelength of the optical signal, which equals to 1550nm in our experiment, and $B_{ele}$ is the bandwidth of the PD, respectively. Note that $\overline{n_M}$ is essentially the sum of all the average photon numbers of every mode. Here, we assume that every mode is of an equal average photon number [48]. Then, the average photon number of every mode can be calculated as

$$\bar{n} = \frac{\overline{n_M}}{M} = \frac{P \cdot 1/B_{ele}}{h\frac{c}{\lambda} \cdot M} \qquad (5)$$

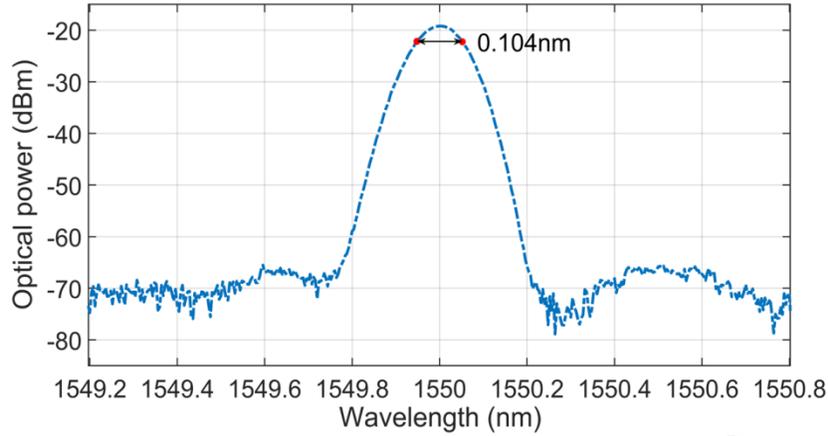

Fig. 3. The optical spectrum of the minimum optical filtering.

For instance, in our experiment, the minimum optical filtering is shown in Fig.3, where a Gaussian shape optical spectrum with a FWHM of 0.104nm ($B_{opt} = 13GHz$) is obtained. The corresponding optical power measured is $P = 33uW$. Then, we can obtain that $M = 2.9627$ and $\bar{n} = 17383$ with equation (3) and equation (5), respectively. The experiments are also performed under several different setups to achieve a comprehensive analysis, which is listed in Table. 1.

**Table 1. Different setups of our experiments.**

| Optical Bandwidth (GHz) | Optical Power (uW) | Mode Number | Average Photon Number |
|---|---|---|---|
| 13 | 33 | 2.9627 | 17383 |
| 16 | 45.4 | 3.5535 | 19939 |
| 23 | 73 | 4.9420 | 23052 |
| 48.5 | 161 | 10.0291 | 25831 |
| 251 | 825 | 50.5203 | 26257 |
| 498.5 | 1660 | 100.0193 | 26681 |

As long as $M$ and $\bar{n}$ are obtained, the theoretical distribution $P_{BE}(n,\bar{n},M)$ of the ASE noise detected in practical experiments can be theoretically calculated, which is purely a function of photon number $n$ and be denoted as $P(n)$. Then the verifications between the theoretical distribution and the experimentally acquired data can be performed. In the following, without the loss of generality, the experiment of minimum optical filtering is used as an instance to introduce our analysis, which also can be applied to others experiment setups.

### 2.3 Physical Model and Numerical Simulation

In the experiment of minimum optical filtering, it is obtained that $M = 2.9627$ and $\bar{n} = 17383$, respectively. Then the theoretical probability distribution $P(n)$ of photon number $n$ can be easily calculated with equation (2), which is shown in Fig. 4.

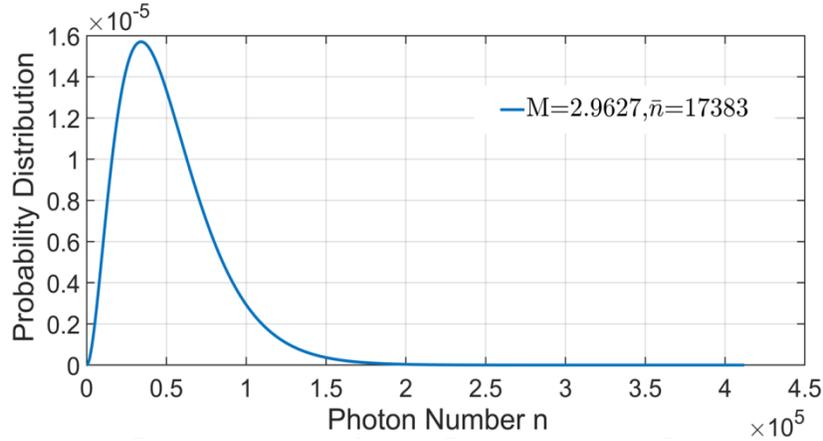

Fig. 4. The theoretical probability distribution of detected photon number calculated when $M = 2.9627$ and $\bar{n} = 17383$, respectively.

It should be noted that the theoretical distribution we obtained is purely the probability of different photon numbers detected during the average detection time of PD with no electronic noise included. However, the experimentally acquired data is essentially a series of voltages that are in principle proportional to the photon numbers detected practically with electronic noise. Therefore, the theoretical distribution is not yet directly comparable with the experimentally acquired data. Thus, to strictly verify the agreement between the theoretical distribution and experimental data, the comprehensive result that comprises both the contributions brought by the photons that obey the theoretical distribution and the electronic noise should be simulated.

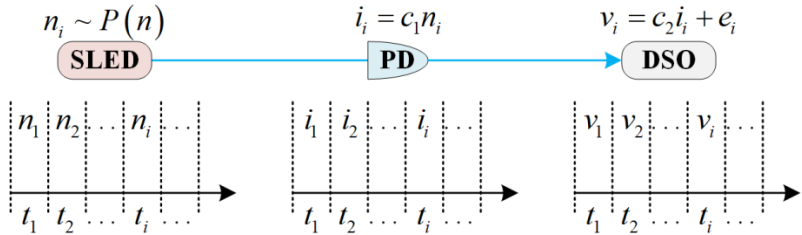

Fig. 5. The physical model for the emission, detection and acquisition of the ASE noise.

To further illustrate our idea, the physical model for the emission, detection and acquisition of the ASE noise is established, which is shown in Fig. 5. We split the whole physical process into a series of procedures, each of which occurs in a time window with a fixed duration equal to the average detection time of PD. In every time window $t_i$, the SLED emits $n_i$ photons, which is an independently identically distribution (i.i.d) random variable that obeys the theoretical distribution $P_{BE}(n,\bar{n},M)$. Then, the PD detects the photons and generates a photon-current $i_i$, which is in principle proportional to the photon number $n_i$. Finally, the DSO acquires the photon-current and obtains a voltage $v_i$, which is proportional to the photon-current $i_i$ with added electronic noise $e_i$. Note that here we suppose the PD is noiseless and ascribe all the electronic noise to a random variable $e_i$, which is in principle independent from $i_i$.

Based on the above physical model, we shall take 3 steps to perform the comprehensive simulation:

***Step1***: We generate a random variable that obeys the theoretical distribution $P(n)$.

This task, in essence, can be ascribed to a general mathematical problem of how to generate the random variable with a given distribution and can be properly solved by employing the Inverse Transform Method, which is described as follows:

*Proposition 1 (The Inverse Transform Method):* Let $F(x), x \in R$ denote any cumulative distribution function (CDF, continuous or not). Let $F^{-1}(y), y \in [0,1]$ denote the inverse function of $F(x)$. Define $X = F^{-1}(U)$, where $U$ is a random variable that obeys uniform distribution over the interval $(0,1)$, i.e. $U \sim Uniform(0,1)$. Then we obtain the random variable $X$ that is distributed as $F$, which equally means that $P(X \leq x) = F(x), x \in R$.

Correspondingly, we first obtain the CDF function of the theoretical distribution by simply calculating

$$F(k) = \sum_{n=0}^{k} P(n), k = 0,1,2,... \qquad (5)$$

Then, we generate the pseudo random variable $U$ with $U \sim Uniform(0,1)$.

Finally, we calculate the expected random variable $N = F^{-1}(U)$ by employing the interpolation method. Thus, $N$ is the random variable that obeys the theoretical distribution $P(n)$. In our simulation, the length of the random variable we generate is $10^7$.

For instance, for the theoretical distribution in the minimum optical filtering setup, the random variable generated by simulation with the Inverse Transform Method is shown in Fig. 6. As expected, the distribution of the random variable generated by simulation fits remarkably well with the theoretical distribution, which in turn proves the validation of the Inverse Transform Method.

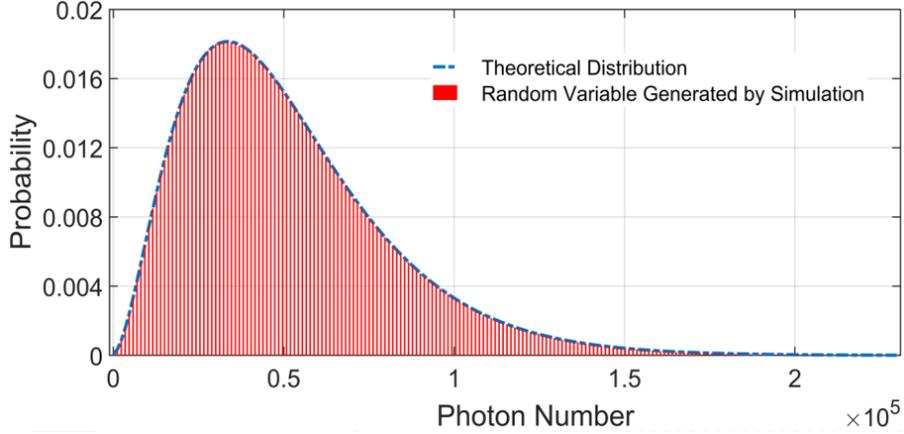

Fig. 6. The distribution of the random variable generated by simulation and the corresponding theoretical distribution for the experiment of minimum optical filtering. The length of the random variable is $10^7$.

***Step2***: We explore the mapping function between the photon number detected in each time window and the corresponding voltage acquired by the DSO.

In Fig.5, it is shown that in principle the numerical expression between the photon number detected in each time window and the corresponding acquired voltage can be descried as

$$v = c \cdot n + e \quad (6)$$

where $v$, $n$ and $e$ denotes the DSO acquired voltage, the detected photon number and the electronic noise, respectively. Intuitively, as long as the coefficient $c$ is achieved, the expected mapping function is also obtained. Thus, in order to calculate the coefficient, another experiment is performed, as shown in Fig. 7.

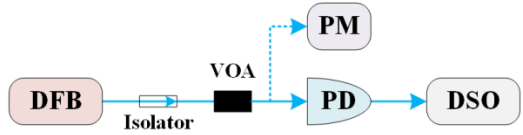

Fig. 7. The experiment setup to explore the mapping function between the photon number detected in each time window and the corresponding voltage acquired by the DSO. A single frequency DFB laser diode is employed to generate optical signal with highly stable initial optical power. A VOA is employed to obtain optical signal with expected optical powers, which is measured by the PM quantitatively and can be used to calculate the corresponding average photon number. Then the attenuated optical signal is detected by PD and finally acquired by DSO to obtain the experimental data. VOA: variable optical attenuator, PD: photodetector, OSA: optical spectrum analyzer, PM: power meter, DSO: digital storage oscilloscope

We first use a single frequency DFB laser diode (Thorlabs, SFL1550P, typical linewidth: 50kHz, center wavelength: 1550nm) with a driving current of 60mA and the initial optical power is 3.86mW. The initial optical signal is attenuated by a VOA to achieve different optical powers. As long as the optical power is stable, the photon number $n$ emitted by the laser in each time window should be approximately constant, which can be measured and calculated with equation (4). Once an expected optical power is obtained, the optical signal is then detected by the PD and the corresponding voltage $v$ can be acquired by the DSO. Note that the acquired voltage still includes electronic noise, so it is reasonable to sample more voltage data and calculate the average value $\bar{v}$ to smooth the electronic noise.

For instance, the minimum optical power we achieved by using the VOA is $P = 30.2uW$ and with equation (4) we can calculate that the photon number in each time window is

$n = 47130$. Then $10^7$ voltage samples are acquired by the DSO and the average value we calculated is $\bar{v} = 1.3963 \times 10^{-3} V$. To strictly validate the mapping function, the experiments are performed under various optical powers, as shown in Table. 2.

Table 2. Different experimental setups to validate the mapping function

| Optical Power (uW) | Photon Number | Average voltage (V) |
|---|---|---|
| 30.2 | 47130 | $1.3963 \times 10^{-3}$ |
| 50.3 | 78498 | $2.3369 \times 10^{-3}$ |
| 100.5 | 156840 | $4.6824 \times 10^{-3}$ |
| 501 | 781857 | $23.201 \times 10^{-3}$ |
| 1000 | 1560593 | $46.256 \times 10^{-3}$ |
| 1507 | 2351813 | $69.857 \times 10^{-3}$ |
| 2005 | 3128989 | $93.074 \times 10^{-3}$ |

Note that, from Table 2 we can see that the optical powers employed for the experiments of SLED (as listed in Table 1) has all been covered, which means that the mapping function obtained from the data in Table 2 is completely applicable to the experimental data of SLED.

We fit the numerical relation of photon number and the average voltage with a linear function, and the following expression is obtained

$$\bar{v} = g(n) = 2.968 \times 10^{-8} \cdot n \tag{7}$$

The agreement between the fit curve and the experimentally acquired data is shown in Fig. 8. A significant agreement between the experimentally acquired data and the linear fit curve is achieved, which effectively demonstrates the validation of the obtained mapping function.

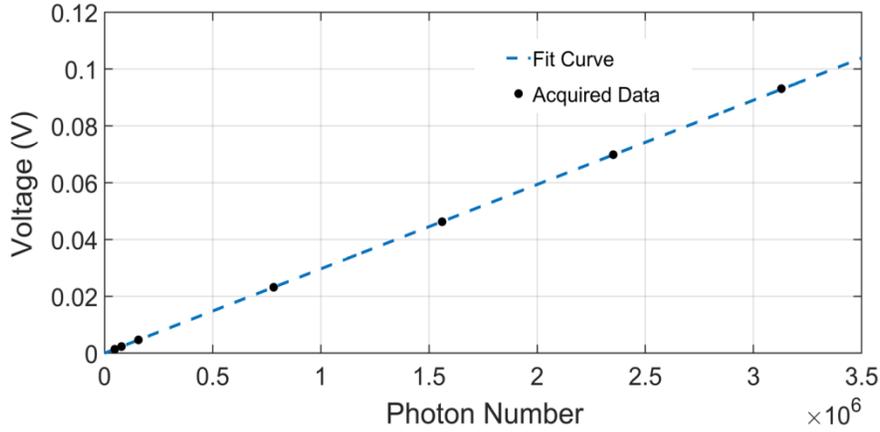

Fig. 8. The fit curve and the experimentally acquired data.

***Step3***: We obtain the comprehensive simulation result.

With the random variable $N$ obtained in step1 and the mapping function $g(n)$ obtained in step2, we first obtain the contributions that are purely resulted from the detected photons by simply calculating

$$V_{photon} = g(N) = 2.968 \times 10^{-8} \cdot N \tag{8}$$

Then, we take the electronic noise into consideration by calculating

$$V_{com} = V_{photon} + V_{ele} \qquad (9)$$

where $V_{ele}$ is the electronic noise acquired previously by turning off the SLED and EDFA.

Now, we consider that $V_{com}$ is the comprehensive simulation result that includes both the contributions brought by the detected photons that obey the theoretical distribution in equation (2) and the electronic noise simultaneously, which is therefore comparable with the previously acquired experimental data $V_{det}$.

## 2.4 Simulation results and Experimental data

Based on the procedures presented in the last subsection, we calculate the comprehensive simulation results under each of the experiment setups presented in Table 1, which includes the scenarios of the mode number $M$ ranging from a relatively small value 2.9627 to a relatively large value 100.0193. The distributions of the comprehensive simulation results and the corresponding experimental data are shown in Fig. 9.

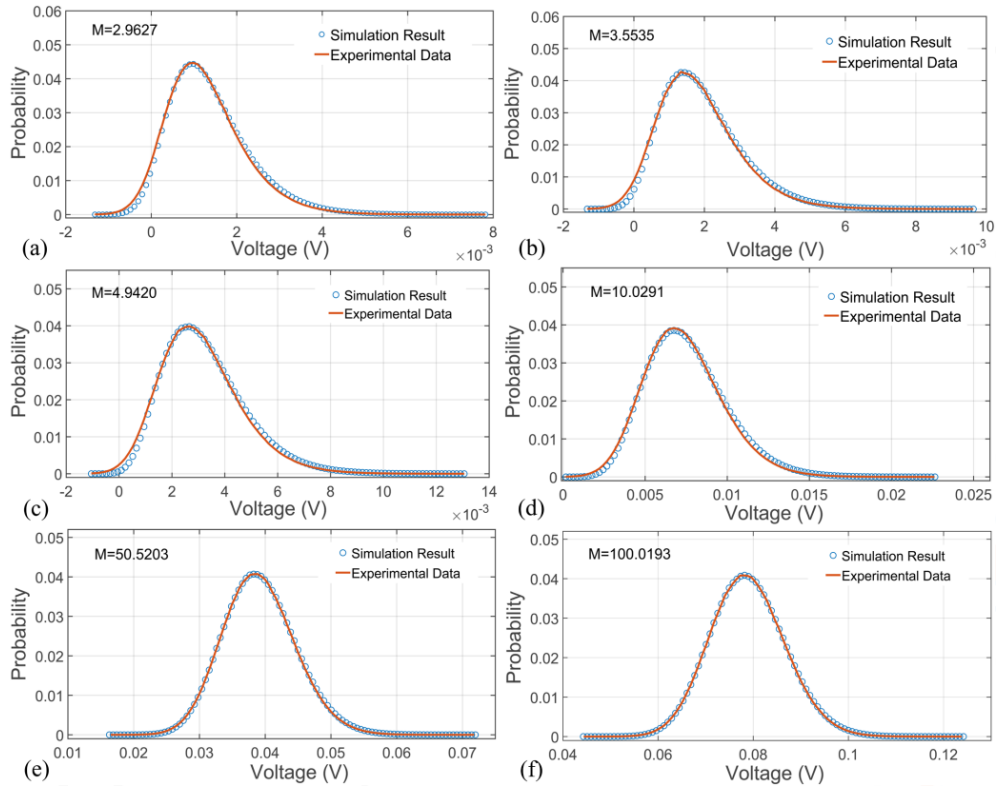

Fig. 9. The distributions of the comprehensive simulation results and the corresponding experimental data under various setups presented in Table 1. (a) Mode number $M = 2.9627$, (b) Mode number $M = 3.5535$, (c) Mode number $M = 4.9420$, (d) Mode number $M = 10.0291$, (e) Mode number $M = 50.5203$ and (e) Mode number $M = 100.0193$.

From Fig.9 it is shown that under each experiment setup, a significant agreement between the comprehensive simulation result and the experimental data is achieved, verifying the validation of the physical model presented in Fig. 5 and the corresponding numerical simulation. The simulation results based on the proposed physical model significantly validate that the photon number distribution of the ASE noise generated by a SLED essentially obeys the *M*-fold degenerate Bose-Einstein distribution, which reveals the randomness origin of the QRNG base on ASE noise.

Note that with the increase of the mode number, the agreement becomes even better. This is probably because that with the increase of the mode number, the optical power and the overall amplitude also increase and therefore the influence on the distribution of the simulation result brought by the electronic noise gradually decays and finally becomes negligible. Thus, the physical model should be more accurate for the detection of the ASE noise with a large mode number.

Specially, we fit the experimental data acquired when the mode number $M = 100.0193$ with a Gaussian distribution function, as shown in Fig.10. It is observed that the Gaussian distribution function fits well with the experimental data. This illustrates that when the mode number is relatively large (e.g. $M > 100$), the detected result of the ASE noise can be practically treated as a Gaussian distribution random variable. Note that in practical QRNGs based on the detection of the ASE noise, the optical bandwidth (usually in the order of THz) is generally much larger than the electrical bandwidth (usually in the order of GHz), which means that the practically detected ASE noise comprises even up to thousands of independent modes and therefore detected results can be treated as a Gaussian random variable.

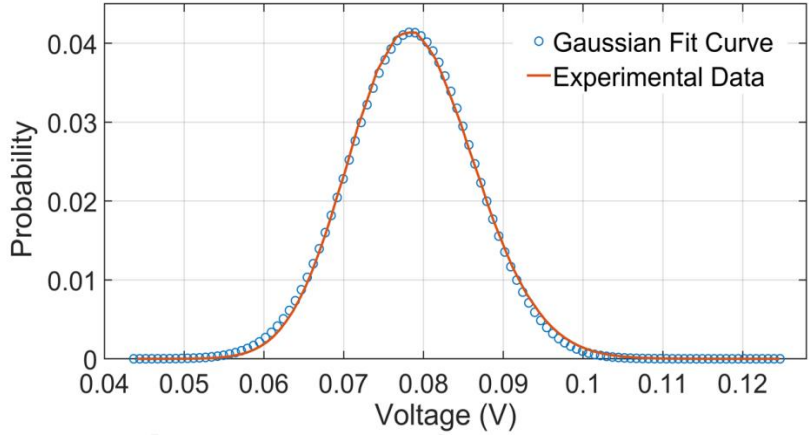

Fig. 10. The distribution of the experimental data with mode number $M = 100.0193$ and the fit Gaussian distribution function.

## 3 Randomness Quantification

In this section, based on the physical model presented in Fig.5, a randomness quantification method and the corresponding experimentally verifiable approach are introduced. In this method, with the theoretical distribution of the detected photon number, the randomness purely resulted from the quantum process is quantified and verified quantitatively by the experimental data, which effectively validates our proposal.

Firstly, we explore how to quantify the randomness based on the theoretical distribution. As stated in [46], if no electronic noise is added and the detection and acquisition system is of arbitrary high precision that every photon number can be resolved, then the randomness that can be extracted is determined by the minimum-entropy, which can be calculated as

$$H_{\min}^{\text{theo}} = -\log_2 \left( \max P_{BE}(n, \bar{n}, M) \right) \quad (10)$$

For instance, in our experiment of the minimum optical filtering, the maximum probability of the theoretical distribution is $P_{\max} = 1.5709 \times 10^{-5}$ and the corresponding minimum-entropy is $H_{\min}^{\text{theo}} = -\log_2 (1.5709 \times 10^{-5}) = 15.9580$. However, in our experiment, the requirement of resolving every photon number is not satisfied and the minimum-entropy calculated with the corresponding experimental data is $H_{\min}^{\text{exp}} = -\log_2 (7.4456 \times 10^{-4}) = 10.3913$, which is obviously smaller than $H_{min}^{theo}$. Therefore, it is necessary to figure out how the limited precision of the detection and acquisition system practically influences the extractable quantum randomness quantified by the minimum-entropy.

Generally, a resolution for a practical detection and acquisition system can be defined, which denotes the minimum increment of the system input that is needed to result in a change of the system output. In our experiment, the resolution stands for the minimum increment of the number of detected photons needed to result in the change of the voltage value acquired on the DSO. Now suppose that the resolution in our experiment is $m$, then this indicates that in principle for the input photon number in the interval of $[i \cdot m, (i+1) \cdot m - 1]$ $i = 0, 1, 2...$, a unique voltage value $v_i$ will be acquired on the DSO, as shown in Fig. 11.

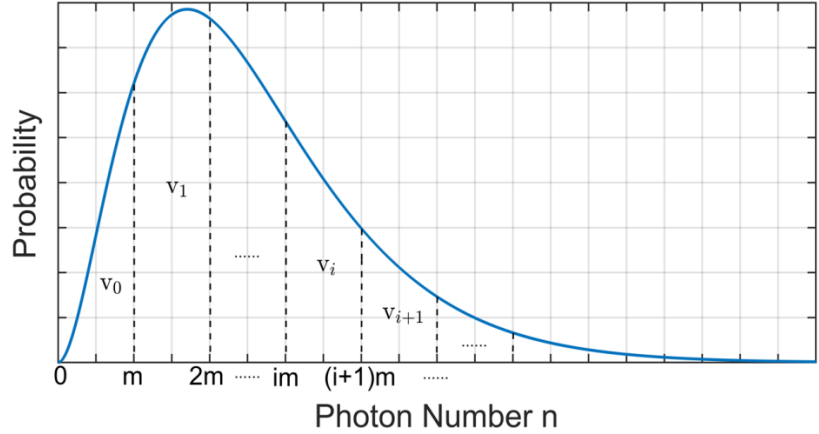

Fig. 11. The acquired voltage with resolution $m$.

From Fig. 11, it can be observed that the corresponding probability of acquired voltage $v_i$ can be obtained by summing every $m$ neighboring values of $P(n)$

$$P'(v_i) = \sum_{k=im}^{(i+1)m-1} P(n) \tag{11}$$

where $P(n)$ is the theoretical distribution of the detected photon number as in Ref. [50]. Specially, when the system resolution is $m=1$, $P'(v)$ should be in principal the same as $P(n)$. However, in practical detection, the system resolution is generally $m \gg 1$, therefore the maximum value of $P'(v)$ is larger than that of $P(n)$ and consequently the corresponding minimum-entropy is smaller, which explains the gap between $H_{min}^{theo}$ and $H_{min}^{exp}$ in our experiment.

Now let's take a deep insight. From equation (11), it is shown that $P'(v)$ can be physically treated as a rearrangement of $P(n)$ by merging its probability values in each interval with a length of $m$. Then, it is feasible to calculate $P'(v)$ quantitatively based on $P(n)$ as long as an approach to achieve the system resolution can be found, which is to be introduced in the following.

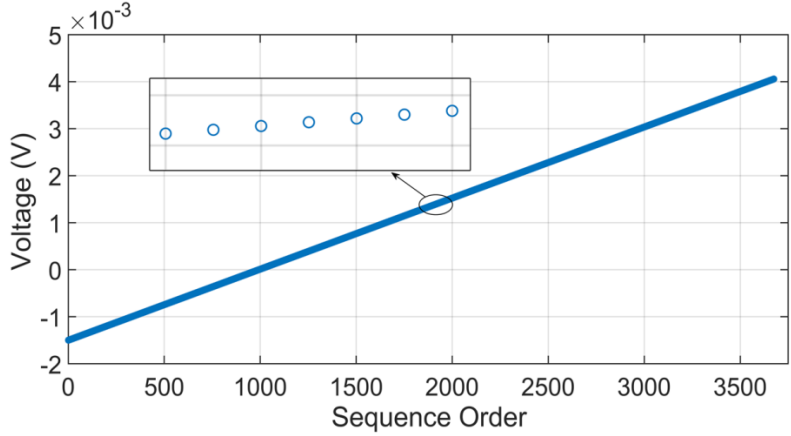

Fig. 12. The unique values of the voltage sample acquired in the experiment.

As in equation (7), the DSO acquired voltage should vary linearly with the photon number detected in each time window and the expected increment resulted by a single photon is $\Delta v_0 = g(n+1) - g(n) = 2.968 \times 10^{-8} V$. Correspondingly, the increment resulted by $m$ photons is $m\Delta v_0$. Here it is supposed that the DSO acquired voltage should in principal vary linearly with a minimum increment of $m\Delta v_0$. Therefore, to verify our hypothesis, we sort the unique values $V_{uni}$ of the voltage sample $V_{det}$ acquired in the experiment of minimum optical filtering in ascending order, as shown in Fig.12. It is illustrated that the DSO acquired voltage indeed varies linearly. Thus, intuitively, the minimum increment of $V_{uni}$ can be obtained by calculating the difference between 2 neighboring unique values. But note that the electronic noise is included in $V_{uni}$. Therefore in order to smooth the electronic noise, we calculate the expected value of the differences of every 2 neighboring unique values and the result is $\overline{\Delta v_{uni}} = 1.5114 \times 10^{-6} V$. Then the system resolution can be calculated as

$$m = \left\lceil \frac{\overline{\Delta v_{uni}}}{\Delta v_0} \right\rceil = \left\lceil \frac{1.5114 \times 10^{-6} V}{2.968 \times 10^{-8} V} \right\rceil = 51 \tag{12}$$

According to equation (11), the probability distribution of the acquired voltage $P'(v)$ can be achieved based on the theoretical distribution $P(n)$ with $m = 51$. As a verification for our approach, the minimum-entropy calculated according to $P'(v)$ is $H_{min}^{merge} = 10.2859$, which is only a deviation of -1.0143% from $H_{min\_exp} = 10.3913$. Since the sampling rate in our experiment is 10G Sa/s, thus by employing proper post processing method (such as the Toeplitz matrix algorithm), an equivalent off-line random number generation rate of 102.859G bits/s can be obtained. Similar calculations are also performed under other setups presented in Table 1 and the results are listed in Table 3, which further validates our proposal.

Table 3. The minimum-entropy calculated with $P'(v)$ under each experiment setup

| Mode Number $M$ | Resolution $m$ | $H_{min}^{merge}$ | $H_{min}^{exp}$ | Deviation | Equivalent Random Number Generation Rate (off-line) |
|---|---|---|---|---|---|
| 2.9627 | 51 | 10.2859 | 10.3913 | -1.0143% | 102.859G bits/s |
| 3.5535 | 51 | 10.6597 | 10.7234 | -0.5940% | 106.597G bits/s |
| 4.9420 | 54 | 11.0834 | 11.1139 | -0.2744% | 110.834G bits/s |
| 10.0291 | 91 | 11.0754 | 11.0848 | -0.0848% | 110.754G bits/s |
| 50.5203 | 109 | 12.0554 | 12.0618 | -0.0531% | 120.554G bits/s |
| 100.0193 | 182 | 11.8375 | 11.8414 | -0.0329% | 118.375G bits/s |

As presented in Table 3, under each experiment setup, the minimum-entropy $H_{\min}^{merge}$ calculated with $P'(v)$ that is obtained through our approach only slightly deviates from the value calculated directly with the experimentally acquired data, which is resulted from the inevitable electronic noise and the statistical fluctuation due to the finite sample size. Therefore, the randomness quantification method and the approach to achieve the system resolution are verified experimentally.

By employing the randomness quantification method presented above, the randomness purely contributed by the quantum process can be obtained, which is of significant importance since it directly indicates how much secure randomness can be extracted and determines the ratio in the post processing for QRNG implementation. Note that in each of our experiments, the minimum-entropy calculated with our method is always smaller than the value directly calculated with the experimental data. This indicates that a smaller but more secure value for the minimum-entropy can be obtained, which is of particular interests for the security analysis of QRNGs. Further note that with the increase of the mode number, the deviation between $H_{\min}^{merge}$ and $H_{\min}^{exp}$ keeps decreasing monotonically. This is also because that the influence brought by the electronic noise gradually decays with the increase of the overall amplitude of the experimental data. As previously mentioned, the detected ASE noise in practical QRNGs comprises up to thousands of independent modes and hence the deviation should be even smaller.

Finally, the calculation method and experimental approach proposed in this section are of significant feasibility and applicable for the QRNG system with randomness originating from the detection of the photon number with arbitrary distributions. For instance, in the QRNG schemes based on the detection of vacuum fluctuations or quantum phase fluctuations, as long as the quantum states for the detected signal can be obtained, the quantum entropy can also be calculated and verified with our proposal.

## 4. Conclusion

In this paper, a systematical physical model for the emission, detection and acquisition of the ASE noise with added electronic noise is developed in detail. By means of experimental validation combined with numerical simulation, the physical model is validated and the statistical distribution of the ASE noise from a SLED in general detection is quantitatively verified. The numerical simulations for the physical model are performed under various setups and the simulation results all significantly fit well with the corresponding experiment data. Then based on the physical model, a randomness quantification method and the corresponding experimentally verifiable approach are proposed and validated, which quantifies the randomness purely resulted from the quantum process and improves the security analysis for the QRNG based on the detection of the ASE noise. The physical model and the randomness quantification method proposed in this paper are of significant feasibility and applicable for the QRNG system with randomness originating from the detection of the photon number with arbitrary distributions.